\documentclass[12pt,preprint]{aastex}

\newcommand{\myemail}{caochen@bao.ac.cn}

\slugcomment{Last Modified Dec 16, 2006 -- Revised Version 3}

\shorttitle{Multi-$\lambda$ Study of YSCs in ARP~24}
\shortauthors{Cao \& Wu}

\begin{document}

\title{Multi-wavelength Study of Young Massive Star Clusters in the Interacting Galaxy ARP~24}
\author{Chen Cao\altaffilmark{1,2}, Hong Wu\altaffilmark{1}}

\altaffiltext{1}{National Astronomical Observatories, Chinese Academy of Sciences, 
Beijing 100012, P. R. China; \myemail; hwu@bao.ac.cn}
\altaffiltext{2}{Graduate School, Chinese Academy of Sciences, Beijing 100039, China}

\begin{abstract}
We made a multi-wavelength study of young massive star clusters (YSCs) in the interacting 
galaxy ARP~24, using the optical and ultraviolet images from {\it Hubble Space Telescope} ({\it HST}), 
Sloan Digital Sky Survey, and {\it Galaxy Evolution Explorer}; the mid-infrared images from {\it Spitzer 
Space Telescope}; and the narrow-band H$\alpha$ image and optical spectra from the NAOC 2.16 m telescope. 
Based on the {\it HST} images, we found that the brightest infrared knot in ARP~24 is associated 
with a complex of five young massive star clusters, within a region of $\sim$ 0.95$''$ radius (127pc) 
in size. The ages and masses of the star clusters in this complex and other regions were estimated 
using {\it HST} broadband photometries and the Starburst99 synthesis models. The star clusters in 
this complex are very young (within ages of $\sim$ 3-5 Myr) and massive (masses of $\sim$10$^{5}$M$_{\rm \odot}$). 
The ionization parameter and metallicity of the complex were estimated using the emission line ratios, 
and the star formation rates were calculated using monochromatic 24$\mu$m, FUV, and H$\alpha$ line luminosities. 
We speculate that ARP~24 may formed by a retrograde fly-by encounter indicated by its one-armed appearance 
and fan-like structure, and the formation of the YSCs in this galaxy is triggered by the interaction. The 
clusters in the YSC complex may formed in a single giant molecular cloud simultaneously. From the ultraviolet 
to mid-infrared spectral energy distributions, we found that the region of the YSC complex is 
relatively bluer in optical and has higher 24$\mu$m dust emission relative to the starlight and 8$\mu$m 
emission. This warm infrared color may due to strong UV radiation field or other mechanisms (e.g., shocks) 
within this region which may destroy the Polycyclic Aromatic Hydrocarbons and enhance the small grain emission 
at 24$\mu$m.
\end{abstract}


\section{Introduction}
Star formation in galaxies generally occurs in star clusters instead of isolated stars, at least 20$\%$ 
and possibly all stars form in clusters or associations (Fall 2004). Young massive star clusters (YSCs, 
with masses often $>$ 10$^{5}$M$_{\rm \odot}$), which are thought to be the products of violent star-forming 
episodes triggered by galaxy collisions, mergers, and close encounters (de Grijs 2003,2004, and references 
therein), or generally form in the disks of isolated spirals with higher efficiency in environments of 
high star formation rate (Larsen 2004a, and references therein), are important for studying the ongoing 
star formation, stellar populations, and the evolutionary histories of their parent galaxies. YSCs are 
thought to formed in Giant Molecular Clouds (GMCs), and concentrated in star-forming clumps (over-dense 
regions, or cores; e.g., Elmegreen 2004).

The majority of important studies of extragalactic star clusters in the last years have involved use of 
the {\it Hubble Space Telescope} ({\it HST}), with its unprecedented spatial resolution ($\sim$0.04$''$) 
and full UV/optical bands (0.1-1.0$\mu$m) access for studying stellar populations, in particular the blue 
coverage for age-dating young clusters (e.g., de Grijs et al. 2003; Larsen 2004b). The ages and masses of 
star clusters can be estimated using color-magnitude and/or color-color diagrams and compared 
with stellar population synthesis models (e.g., Johnson et al. 2003; Harris et al. 2004). 
Previous studies of extragalactic star clusters paid more attention to their general properties based 
on optical images, such as their luminosity and mass functions (e.g., Whitmore et al. 1999; Elmegreen 
et al. 2001), the size of the clusters (e.g., Larsen 1999), and the mechanisms of cluster formation and 
disruption (e.g., Elmegreen 2004). 

Nevertheless, mid-infrared observations of extragalactic star clusters are also very important, for 
understanding the properties of heavily obscured clusters (e.g., Bontemps et al. 2001), and dust environments 
of young cluster-forming systems (e.g., Zhang et al. 2001). The {\it Spitzer Space Telescope}'s (Werner et al. 2004) 
observations in mid-infrared with higher sensitivity and better angular resolution than previous observations 
(e.g., {\it ISO}), provide a new opportunity to study both young and old stellar populations and star formation 
in normal (e.g., Pahre et al 2004; Wu et al. 2005; Calzetti et al. 2005), starburst (Cannon et al. 2005,2006ab), 
and interacting/merging (e.g., Wang et al. 2004; Smith et al. 2005; Elmegreen et al. 2006) galaxies. The four 
IRAC bands from 3.6 to 8.0 $\mu$m probe both stellar continuum and warm dust emissions (of the so-called Polycyclic 
Aromatic Hydrocarbon, or PAH, and dust continuum emissions), and MIPS 24$\mu$m band probes the warm dust emissions 
from the Very Small Grains (VSGs). Although {\it Spitzer} images are unable to resolve individual star clusters 
which can be well resolved by {\it HST}, they can be used to study the mid-infrared properties of the infrared 
bright knots/clumps which may be collections of OB associations and dense clusters of several hundred parsecs 
in size (Efremov 1995; Elmegreen et al. 2006).  

ARP~24 (NGC~3445) forms a triplet with NGC~3440 and NGC~3458, at a separation of 9.9$'$ and 14.0$'$, respectively. 
Numerous H {\sc ii} region candidates exist in its spiral arm and disk. At the end of its southern spiral arm is 
a shred that may be a separate galaxy or was one in the past. ARP~24 is at distance of 27.6 Mpc (1$''$ corresponds 
to 134 parsecs) and with a total infrared luminosity of L$_{\rm \rm TIR}$ $\sim$ 4.8$\times$10$^{9}$L$_{\rm \rm \odot}$ 
(Bell 2003). Van den Bergh (1995) classified it as a 'transitional object' that appear intermediate between spirals 
that have central bulges and objects having central regions that are resolved into stars and knots. B{\"o}ker 
et al. (2002,2003) found it contains a nuclear star cluster in the central region instead of a spiral bulge. ARP~24 
is a target of the ``Spiral, Bridges, and Tails'' (SB\&T) Guest Observer Cycle 1 {\it Spitzer} program (PI: C. Struck), 
which is for studying the distribution of star formation in a sample of colliding galaxies with a wide range of tidal 
and splash structures (see Smith et al. 2005,2006; Hancock et al. 2006 for details).

In this paper, we present an analysis of data from {\it HST}, {\it Spitzer}, SDSS, {\it GALEX}, 
H$\alpha$ image, and spectroscopic observations, for studying the properties and possible formation 
scenarios of the YSCs in the interacting galaxy ARP~24. The observations and relevant data reduction 
are presented in $\S$2; results on the multi-wavelength emissions from the YSCs, the discovery of a 
YSC complex (YSCC), and other physical properties of the YSCs are described in $\S$3. Possible formation 
scenarios of the YSCs, and the PAH and warm dust emissions of the infrared knots in this system are 
discussed in $\S$4. The major results of this work are summarized in $\S$5.

\section{Observations and Data Reduction}

\subsection{Optical and Ultraviolet Images}
The U (F300W) and I-band (F814W) images of ARP~24 were taken with the WFPC2 Wide Field Camera and 
Planetary Camera on board {\it HST}, with exposure times of 600 and 640 seconds, respectively. The 
images were obtained as B associations from the ESO/ST-ECF Science Archive
$\footnote{http://archive.eso.org/wdb/wdb/hst/science/form}$
. The 'PHOTFLAM' (in ergs$^{-1}$cm$^{-2}$\AA$^{-1}$), 'EXPTIME' (exposure duration) keywords in the 
header of each image were used for converting instrumental magnitudes to flux densities, and to VEGA 
magnitudes based on the zeropoints given in WFPC2 data handbook
$\footnote{http://www.stsci.edu/instruments/wfpc2/Wfpc2$\_$dhb}$
. The broadband optical images ($u,g,r,i,z$) were derived from the SDSS data archive (York et al. 2000; 
Stoughton et al. 2002). The background fitted by low-order Legendre polynomial was subtracted from 
each band image, after masking out all bright sources (Zheng et al. 1999; Wu et al. 2002). Then the 
counts were converted to flux densities and AB magnitudes
$\footnote{http://www.sdss.org/dr5/algorithms/fluxcal.html; see also Fukugita et al. (1996).}$
. Figure 1 shows the three color image of ARP~24 derived from the SDSS data archive. North is up, 
east is to the left as denoted by the cross hair.

The far-ultraviolet (FUV; 1516\AA) and near-ultraviolet (NUV; 2267\AA) images were derived from the ultraviolet 
atlas of nearby galaxies (Gil de Paz et al. 2006) based on images obtained with NASA's satellite, the {\it 
Galaxy Evolution Explorer} ({\it GALEX}; Martin et al. 2005). Keywords of mean sky-background level (SKY) and 
zero point in AB magnitudes scale (ZP) in the headers of the FUV \& NUV images were used for sky-subtractions 
and flux calibrations, respectively. 

The narrow-band interference filter image centered near the red-shifted H$\alpha$ and the associated continuum 
filter (R band) image of ARP~24 were taken on 23 February, 2006, by the 2.16 m telescope at Xinglong observatory 
of NAOC
$\footnote{http://www.xinglong-naoc.org/English/216.html}$
, with the BAO Faint Object Spectrograph and Camera (BFOSC), using a LICK 2048 $\times$ 2048 CCD as detector. 
The field of view of the CCD is close to 10$'$ $\times$ 10$'$, with a pixel size of 0.305$''$. The ``H$\alpha$'' 
images actually include a component from the [NII]$\lambda$$\lambda$6548,6583 doublet in addition to the Balmer 
line flux. Standard CCD reductions include overscan and bias subtractions, flatfield correction and cosmic-ray 
removal were applied to the images using IRAF package. Then the astrometric calibrations were computed to place 
the images on the SDSS frames, using several field stars as reference, and the accuracy of the calibration is 
better than 1$''$. SDSS images were used as references to our R-band image for flux calibration, based on 
field stars and the transformations between SDSS magnitudes and UBVRcIc given by Lupton (2005)
$\footnote{http://www.sdss.org/dr5/algorithms/sdssUBVRITransform.html$\#$Lupton2005}$
. R-band image was used for subtracting the stellar continuum from the H$\alpha$ image. We selected five unsaturated 
stars over the field of view and adjusted values from the scaled R-band subtracted H$\alpha$ image that these stars 
were canceled in a statistical sense.

\subsection{Infrared Images}
Broadband mid-infrared images of ARP~24 were acquired with the Infrared Array Camera (IRAC, Fazio et al. 2004) 
and Multiband Imaging Photometer for Spitzer (MIPS, Rieke et al. 2004) on board the {\it Spitzer Space Telescope}. 
The Basic Calibrated Data (BCD) were part of the Lockman Hole field in the {\it Spitzer} Wide-field Infrared 
Extragalactic (SWIRE) Survey (Lonsdale et al. 2003). Following preliminary data reduction by the {\it Spitzer} 
Science Center pipeline, images of each of the four IRAC (3.6, 4.5, 5.8 and 8 $\mu$m) and MIPS 24$\mu$m bands 
were mosaicked, after pointing refinement, distortion correction and cosmic-ray removal (Fazio et al. 2004; 
Huang et al. 2004; Wu et al. 2005; Surace et al. 2005). The mosaicked images have pixel sizes of 0.6$''$ and 
angular resolutions (full width at half maximum, FWHM) of 1.9$''$, 2.0$''$, 1.9$''$ and 2.2$''$ for IRAC four 
bands, and pixel size of 1.225$''$ and FWHM of 5.9$''$ for MIPS 24$\mu$m band, respectively. Figure 2 shows 
{\it Spitzer} IRAC 3.6, 4.5, 5.8, 8$\mu$m and MIPS 24$\mu$m images of ARP~24. The continuum-subtracted 
H$\alpha$ image was also shown to emphasize the good correspondence of the peaks between H$\alpha$ and 
24$\mu$m emissions.

\subsection{Aperture Photometry}
Aperture photometries were performed on {\it HST} WFPC2 F300W \& F814W images using 0.2$''$ radius 
circular apertures. Photometries on SDSS, {\it GALEX}, {\it Spitzer}, and continuum-subtracted 
H$\alpha$ images were performed using apertures of 12$''$ and 120$''$ diameters for individual 
regions and the entire galaxy, respectively. All photometric values were derived without aperture 
corrections. Both the IRAC and MIPS 24$\mu$m bands have calibration uncertainties at the 10$\%$ 
level (Dale et al. 2005; Wu et al. 2005). The global values of the 70$\mu$m and 160$\mu$m fluxes of 
ARP~24 were derived from the catalogues of SWIRE data release 3 (Surace et al. 2005).

\subsection{Optical Spectroscopy}
The optical spectrum of the YSC complex (see $\S$3.2) in ARP~24 was taken on 07 April, 2006. The 
observation was carried out with 2.16 m telescope and the BFOSC. A G4 grism (from 4000 to 8000 \AA) 
and slit width of $\sim$ 2$''$ were used in this observation, gives a dispersion of roughly about 
200\AA/mm. A relatively higher resolution spectrum of this complex was taken on 06 May, 2006, with 
2.16 m telescope and the OMR spectrograph. A 1200 l/mm grating (from 6300 to 7100 \AA), blazed at 6700\AA\/, 
and slit width of 2$''$ were used in this observation, gives a dispersion of about 50\AA/mm. Optical 
spectrum of the nuclear region in ARP~24 was taken on 05 May, 2006, with the OMR spectrograph and a 
300 l/mm grating (from 3800 to 8000 \AA) with slit width of 2$''$. The unprocessed frames were reduced by 
standard CCD procedure using IRAF package. The CCD reductions include bias subtraction, flatfield correction 
and cosmic-ray removal. The wavelength calibrations were carried out using Fe/Ar (for BFOSC) and He/Ne/Ar 
(for OMR) lamps. KPNO IRS standard stars were observed each night for carrying out the flux calibrations.

\section{Results}

\subsection{Multi-wavelength Emission in ARP~24}
From SDSS (Fig. 1) and {\it Spitzer} (Fig. 2) images, we found that the nucleus of ARP~24 appears 
relatively redder in optical and bright at 3.6 and 4.5 $\mu$m, but is absent in IRAC 8$\mu$m and MIPS 
24$\mu$m bands. In the MIPS 24$\mu$m image we identified four infrared bright knots (labeled with K0, 
K1, K2, K3) in ARP~24, which are bluer in optical and bright at 8$\mu$m and 24$\mu$m. The brightest 
infrared knot (K0) was found to be associated with a star cluster complex based on {\it HST} images 
(see next section), so we re-labeled it with 'YSCC'. Many gaseous structures and filamentaries were 
also shown up in IRAC 5.8, 8$\mu$m and MIPS 24$\mu$m bands. Most of the emission at 8$\mu$m (and 
also at 5.8 $\mu$m to a lesser extent) arises from the so-called aromatic features (usually attributed 
to PAHs) at 6.2, 7.7, and 8.6 $\mu$m in the photo-dissociation regions (PDR), which are often associated 
with the warm dust in or near star forming regions (H {\sc ii} regions), thus can be used as tracers of 
star formation (Peeters et al. 2004; Wu et al. 2005). MIPS 24$\mu$m emissions, which is mainly due to hot 
dust emissions from the Very Small Grains (VSGs), were also thought to be good measures of the SFRs of 
galaxies (Wu et al. 2005; Calzetti et al. 2005; P{\'e}rez-Gonz{\'a}lez et al. 2006). Thus, we conclude 
that the four infrared bright knots (YSCC, K1, K2, K3) are dominated by young stars and are sites of 
active star formation. The H$\alpha$ image of ARP~24 shows strong H$\alpha$ emissions in these regions, 
which also indicates young stellar population there (see Fig. 2).

The ultraviolet to infrared Spectral Energy Distributions (SEDs) were shown in Figure 3 and Table 1. From 
comparison with the empirical SED templates generated using the GRASIL code (Silva et al. 1998; see also more 
detailed descriptions in Jarrett et al. 2006), we find that the YSCC, which has very strong 24$\mu$m dust 
emission relative to PAHs, is very similar to the prototype starburst galaxy M~82 in mid-infrared, but exhibits 
large excess of starlight at ultraviolet and short optical wavelength relative to the M~82 template (see Figure 3b). 
This is similar to the tidal tail super massive clusters in Tadpole, in which the significant blue excess is 
considered to be related to recent star formation (Jarrett et al. 2006) and the gas/dust was expected to be 
blown away by stellar winds and/or supernovae explosions. The other three infrared bright knots (K1, K2, K3), 
which indicate strong PAH and dust continuum emissions, are comparable to star formation regions in late-type Sc 
galaxies (e.g., NGC~6946) and the infrared-bright regions of the Tadpole disk (Jarrett et al. 2006).

\subsection{A Young Massive Star Cluster Complex Associated with the Brightest Infrared Knot}
The brightest infrared knot in ARP~24 (K0) was found to be associated with a complex of five star clusters 
('YSCC' for short), based on U and I-band WFPC2 images from {\it HST} (Fig. 4). This complex is within 
a region of $\sim$ 0.95$''$ radius (127pc) in size, and apart from the nuclear star cluster in the galaxy 
center by about 20.5$''$ ($\sim$ 2.7kpc). The measured fluxes and absolute magnitudes of the star clusters 
in the YSCC, the nucleus, and other three infrared bright knots (K1, K2, K3) were shown in Table 2
$\footnote{Part of the southern primary disk and the companion galaxy in the east of ARP~24 are not contained 
in the {\it HST} images. Note our study may suffers some selection effects, because we can only analyze several 
brightest clusters (which may be biased to younger populations) due to the relative shallowness of {\it HST} 
images (especially for the F300W), so our results can not be used for statistical studies of star clusters in 
this system.}$
. The different m$_{\rm F300W}$-m$_{\rm F814W}$ colors between clusters in each region primarily reflect the mean 
age of the stellar population. 

The color-magnitude diagrams (CMD) of the YSCC and three infrared bright knots in ARP~24 are 
shown in Figure 5. The ages and masses of the clusters were estimated using Starburst99 instantaneous 
models (Leitherer et al. 1999), with Salpeter initial mass function (IMF, $\alpha_{\rm IMF} = 2.35$) 
between 0.1 and 120 M$_{\rm \rm \odot}$ and metallicities Z=0.02 \& 0.008. The models are also shown in the CMDs, 
ages along the evolutionary lines which are drawn for cluster masses of 10$^{5}$ (solid) and 4$\times$10$^{4}$ 
(dotted) M$_{\rm \odot}$ are indicated by diamonds (Z=0.02) and plus (Z=0.008) place in t=1 Myr intervals and begin 
from 1 Myr on the left. The clusters in the YSCC and K1 are within ages of around 3-5 Myr and masses of about 
10$^{5}$M$_{\rm \odot}$. The clusters in other two infrared bright regions (K2 \& K3) are relatively less massive 
($\sim$ 4$\times$10$^{4}$M$_{\rm \odot}$) but have similar ages ($\sim$ 4-6 Myr) to that in the YSCC and K1. The 
masses estimated here are consistent with that derived from {\it HST} F300W/F814W luminosities and the 
mass-to-light ratio (M/L) of $\sim$ 0.01-0.02 in visual band for young clusters with a 10$^{6.7}$ burst 
(Chandar et al. 1999). The masses of the star clusters in the YSCC and K1 are consistent with that of YSCs 
or super star clusters (de Grijs 2003) observed in other interacting galaxies (e.g., IC~2163 \& NGC~2207, 
Elmegreen et al. 2001; NGC4038/39, Whitmore et al. 1999), which are thought to be the progenitors of luminous 
globular clusters (e.g., Ma et al. 2006ab; Kravtsov 2006).

\subsection{Optical Spectroscopy}
Optical spectrum of the nucleus in ARP~24 (Fig. 6) indicates a mixture of populations of different ages, 
with strong H$\delta$ absorption which indicates the existence of a large number of evolved A-type stars 
(age $\sim$ 10$^{8}$yr; see, e.g., Wang \& Wei 2006), and strong H$\alpha$ emission which traces ongoing star 
formation. A template spectrum corresponds to an instantaneous-burst model with a young population of 100 Myr 
and metallicity Z=0.02 plus an old population of 11 Gyr was also plotted for comparison. This result is in 
agreement with previous studies that the nuclear clusters are massive and dense star clusters which form stars 
recurrently until the present day (Walcher et al. 2005,2006). Unfortunately, its HST/STIS spectra which can 
provide better separation of nuclear star cluster light from underlying galaxy light than our ground-based 
spectrum can't be used for stellar population analysis (due to its low S/N ratio; $\sim$ 2.9, see Rossa et al. 
2006).

Optical spectrum of the YSCC in ARP~24 (Fig. 7) shows strong hydrogen and oxygen emission lines which indicates 
young stellar populations and active star formation. The derived redshift is 0.007, matching the redshift (0.0069) 
of the galaxy derived from the Updated Zwicky Catalog (Falco et al. 1999). The emission lines were fitted with 
Gaussian profiles and the line fluxes were listed in Table 3. The line ratios are H$\alpha$/H$\beta$=3.13, 
[OIII]/H$\beta$=3.57, and [NII]/H$\alpha$=0.12,[SII]/H$\alpha$=0.19. Using standard optical line ratio diagnostic 
diagrams and comparing with the 'MAPPINGS III' code (Kewley et al. 2001; Kewley \& Dopita 2002) results, we 
computed photo-ionization models and estimated the ionization parameter q (cm s$^{-1}$) of the YSCC, which is 
about 4$\times$10$^{7}$. 

The dust extinction can be estimated from the Balmer decrement (Calzetti 2001):
\begin{displaymath}
E(B-V) = 2.5\frac{log(\frac{H\alpha/H\beta}{H\alpha_{0}/H\beta_{0}})}{k(H\beta)-k(H\alpha)}
\end{displaymath}
where the value of the intrinsic luminosity ratio H$\alpha$$_{\rm 0}$/H$\beta$$_{\rm 0}$ is 2.87 for temperature T = 10,000 
K and case B recombination (Osterbrock 1989), and the value of the differential extinction k(H$\beta$)-k(H$\alpha$) 
between H$\alpha$ and H$\beta$ is 1.163 (Calzetti 2001). Then the extinction of the YSCC can be calculated from 
A$_{\rm V}$=R$_{\rm V}$E(B-V) (R$_{\rm V}$ = 3.1), and the value is about 0.25. It is smaller than the attenuations 
in M~81 (A$_{\rm V}$$\sim$0.5; see, e.g., Hill et al. 1995), M~51 (A$_{\rm V}$$\sim$3; see Calzetti et al. 2005), and 
the archetype starburst galaxy M~82 (A$_{\rm V}$$\sim$0.5; Mayya et al. 2006).

The metallicity was estimated using the line ratios, following Vacca 
\& Conti (1992):
\begin{displaymath}
log(O/H) = -0.69logR_{3} - 3.24  (-0.6 \leq logR_{3} \leq 1.0)
\end{displaymath}
where
\begin{displaymath}
R_{3} = \frac{I([OIII]\lambda4959)+I([OIII]\lambda5007)}{I(H\beta)}
\end{displaymath}
The measured metallicity (12+log(O/H)) of the YSCC is about 8.37, or 0.51Z$_{\rm \odot}$ if adopted the value 
of 8.66 for solar abundance (Asplund et al. 2004). This relation (R$_{\rm 3}$$\sim$log(O/H)) has been shown 
to be affected by the ionizing photon hardness. Thus, we also estimated metallicities using the 
[NII]/H$\alpha$ ratios for a comparison, with:
\begin{displaymath}
12+log(O/H) = 9.12 +0.73log\frac{[NII]}{H\alpha} 
\end{displaymath}
(Kewley \& Dopita 2002). The metallicity of the YSCC estimated using this relation is about 8.45, 
consistent with that derived from the R$_{\rm 3}$ value. The metallicity of the nucleus in ARP~24 was also 
estimated using the [NII]/H$\alpha$ ratio, and the value is about 8.79, much higher than that of the YSCC.

\subsection{Star Formation Rate and Stellar Mass}
The total star formation rate (SFR) of ARP~24 was calculated using infrared luminosity derived from 
{\it Spitzer} MIPS (24, 70, 160$\mu$m) and the equation of Dale \& Helou (2002): 
\begin{displaymath}
L_{TIR} = 1.559 \nu L_{24} + 0.7686 \nu L_{70} + 1.347 \nu L_{160}
\end{displaymath}
Then the SFR$_{\rm total}$ can be estimated by multiplying the L$_{\rm IR}$ (in ergs s$^{-1}$) by a conversion 
factor of 4.5$\times$10$^{-44}$ M$_{\rm \odot}$yr$^{-1}$ (Kennicutt 1998)
$\footnote{Note, different definitions of L$_{\rm IR}$, e.g., L(3-1100$\mu$m) used in this paper by Dale \& Helou 
(2002) and L(8-1000$\mu$m) from Sanders \& Mirabel (1996), will not affect much (with an uncertainty of $\sim$ 15\%; 
see also Moustakas et al. 2006) on estimating SFRs in the ARP~24 system.}$
. The total SFR in ARP~24 is about 1.05M$_{\rm \odot}$yr$^{-1}$. The SFRs of the YSCC and K1, K2, K3 were estimated 
using 24$\mu$m dust emissions, which are thought to be good measures of the SFRs of galaxies (Wu et al. 2005; 
Calzetti et al. 2005; Perez-Gonzalez et al. 2006), and equation (3) of Wu et al. (2005):
\begin{displaymath}
SFR_{\rm 24\mu m}(M_{\rm \odot}yr^{-1}) = \frac{\nu L_{\rm \nu}(24\mu m)}{6.43\times10^{8}L_{\rm \odot}}
\end{displaymath} 
the measured SFRs in the regions of the YSCC, K1, K2, and K3 are about 0.10, 0.03, 0.04, 0.03 
M$_{\rm \rm \odot}$yr$^{-1}$, respectively. The SFR in YSCC is extremely high, considering its relatively small 
spatial size comparing with the entire galaxy. The SFR per unit area of this complex was estimated to be 
0.05 to 1.97 M$_{\rm \rm \odot}$yr$^{-1}$kpc$^{-2}$ using 24$\mu$m 6$''$ radius and {\it HST} 0.95$''$ radius, 
respectively. It is comparable to or even much stronger than the definition of a starburst galaxy ($\sim$0.1 
M$_{\rm \rm \odot}$yr$^{-1}$kpc$^{-2}$, Kennicutt et al. 2005), thus the YSCC can be justified as a localized 
starburst (Efremov 2004). 

Although using infrared emission to estimate SFRs is more indirect than other young star tracers such as 
UV and H$\alpha$ line emissions, it suffers relatively minor extinction effects which are difficult to 
correct. From the simulations of the performance of star-formation indicators in the presence of dust, 
Jonsson (2004) found the infrared luminosity is more reliable than H$\alpha$ and FUV luminosities which suffer 
severely from dust attenuation and the situation can only partially be remedied by dust corrections. For 
comparison we calculated the SFRs based on H$\alpha$ line and FUV luminosities using the relations: 
SFR$_{\rm H\alpha}$(M$_{\rm \odot}$yr$^{-1}$) = 7.9$\times$10$^{-42}$ L$_{\rm H\alpha}$(erg s$^{-1}$) (Kennicutt 1998) and 
log SFR$_{\rm FUV}$(M$_{\rm \odot}$yr$^{-1}$) = log L$_{\rm FUV}$(L$_{\rm \odot}$) - 9.51 (Iglesias-P{\'a}ramo et al. 2006). 
The SFR$_{\rm H\alpha}$ and SFR$_{\rm FUV}$ (no reddening corrected) for each region are about 0.14, 0.03, 0.04, 0.04 
M$_{\rm \rm \odot}$yr$^{-1}$ and 0.11, 0.03, 0.05, 0.04 M$_{\rm \rm \odot}$yr$^{-1}$, respectively, a bit higher than 
that derived from the 24$\mu$m luminosities. A better way for estimating SFRs may be based on a combination of 
the observed infrared and ultraviolet/optical luminosities as suggested by some authors recently (e.g., Kennicutt 
2006; Iglesias-P{\'a}ramo et al. 2006). We adopted the relations: 
SFR(M$_{\rm \odot}$yr$^{-1}$) = 4.5 L$_{\rm TIR}$ + 7.1 L$_{\rm FUV}$ (10$^{37}$W) (Dale et al. 2006) and 
log L(TIR) = log L(24) + 0.908 + 0.793 log [L$_{\rm \nu}$(8)/L$_{\rm \nu}$(24)] (Calzetti et al. 2005), then 
the SFR$_{\rm IR+FUV}$ for each region are about 0.14, 0.05, 0.07, 0.06 M$_{\rm \rm \odot}$yr$^{-1}$, respectively, 
slightly higher than previous estimations. Due to the relatively small variations of SFRs derived from different 
SFR indicators, we adopted the values of SFR$_{\rm 24\mu m}$ for further analysis. 

The stellar masses for the old stellar population in different regions of ARP~24 were estimated based on the 
SDSS photometries following Bell et al. (2003),
\begin{displaymath}
\log (Mass_r/M_{\odot}) = -0.4(M_{\rm r, AB} - 4.67)+[a_{\rm r} + b_{\rm r}\times (g-r)_{\rm AB} + 0.15]
\end{displaymath} 
where the coefficients $a_{\rm r}$ (-0.306) and $b_{\rm r}$ (1.097) are taken from Table 7 of Bell et al. 
(2003). The estimated stellar masses in the regions of the YSCC, K1, K2, and K3 are about 10$^{8.5}$, 
10$^{8.6}$, 10$^{8.5}$, 10$^{8.4}$ M$_{\rm \rm \odot}$, respectively. We also compared the measured stellar 
masses with the monochromatic IRAC 3.6$\mu$m luminosities (which are around 10.57, 8.55, 8.10, 6.81 
$\times$10$^{40}$ergs s$^{-1}$, respectively), which are thought to be approximate measures of the stellar 
masses in galaxies (e.g., Smith et al. 2006). We found the regions in ARP~24 followed a similar scaling 
relation between Mass$_{\rm r(R)}$ and logL(3.6$\mu$m) to that in the ARP~82 system (Hancock et al. 2006). 
The specific star formation rate (SSFR, SFR per unit stellar mass, in units of Gyr$^{-1}$) for each region 
are 0.32, 0.08, 0.14, 0.13 Gyr$^{-1}$, respectively. The SSFRs of these regions in ARP~24 are higher 
than that of most of the local star-forming galaxies (with values between 0.03 and 0.2 Gyr$^{-1}$; Bell 
et al. 2005), but are much lower than that of Luminous Infrared Galaxies (LIRGs) in the local universe 
(lie between 1.2 and 10 Gyr$^{-1}$; Wang et al. 2006). However, the SSFRs estimated here will only be a 
lower limit for the SSFRs within or around the YSCs in themselves, due to the contamination of star lights 
(dominated by old population) from their parent galaxy.

\section{Discussion}

\subsection{Possible Formation Scenarios of the YSCs in ARP~24}
'Peculiar one-time events and special places that have extraordinary high energy inputs' was 
suggested by Elmegreen (2004) as one of the possible mechanisms for triggering the formation of 
YSCs. He suggested that YSCs in galaxies undergoing interactions may formed by large part or 
local shock compressions and collapse from cloud impacts or colliding supershells. Bastian et al. 
(2006a) studied the young star cluster complexes in Antennae and suggested that if we assume that 
the grouping of complexes formed out of the same GMC, then the star formation is triggered by an 
external perturbation. Cannon et al. (2005) studied the infrared properties of the supergiant shell 
region of the dwarf galaxy IC~2574 using {\it Spitzer} and demonstrate that the expanding 
shell is affecting its surroundings by triggering star formation and heating the dust. The morphology 
of ARP~24 indicates that it may formed in a minor merger scenario (interaction between a gas-rich 
late-type spiral and a small companion) or from cosmological accretion of gas on galactic disks 
(Bournaud et al. 2005). Shaping/reshaping of galaxies is the biggest effect generally attribute 
to mergers (Cox 2004). The tidal stripping of a satellite can produce features such as long tails, 
and the accretion introduces new stellar populations into the galactic disk (Walker et al. 1996), 
i.e., promotes a high SFR (e.g., Mihos \& Hernquist 1994,1996; see also Cox 2004) in ARP~24. The YSCs 
may form as a consequence of the high SFR per unit area in the galaxy (Larsen \& Richtler 2000). ARP~24 
shows a strong asymmetry in g-i colors (see Fig. 8) through the galactic disk, the regions near the YSCC 
are relatively bluer than that near the nucleus. This asymmetry may be due to the differences in star formation, 
stellar populations, gas/dust contents of different regions caused by galaxy interactions. Furthermore, 
the one-armed appearance and the broad fan-like structure seen in the primary disk of ARP~24 indicate that 
this galaxy may have formed by a retrograde fly-by encounter (see, e.g., Barnes \& Hernquist 1996; Struck 
\& Smith 2003; and a review by Struck 1999, and references therein). 

The companion galaxy in the east seems nearly edge-on (also a bit warp), with a weak bridge connects to the 
primary disk and faint knots near the outer edges of its disk (see Figs. 1\&2). This could indicates the 
existence of an edge-on ring (or other waves), which would be consistent with a small angle between the orbital 
plane of the companion and the disk plane of the primary, if the closest approach was on the west side. It is 
also possible that the primary disk might have rotated by about half a turn in the time since the closest 
approach. Thus, the region of the YSCC in ARP~24 might be in the part of the disk that was closest to the 
companion and most perturbed at that time. If we adopt a rotation velocity of about 100 km s$^{-1}$ (derived 
from the rotation curve of M~51
$\footnote{Derived from the Fabry-Perot de Nouvelle Technologie pour l'Observatoire du mont M$\acute{e}$gantic 
(FaNTOmM); see Daigle et al. (2006).}$
) and the distance between YSCC and the nuclear star cluster ($\sim$ 2.7kpc) as the radius of rotation, then 
such half a turn will takes about 8$\times$10$^{7}$yrs; and moreover, if we assume a circular orbit of the 
encounter and a relative velocity of $\sim$ 400 km s$^{-1}$ between the companion and the primary's core 
(Eneev et al. 1973), then the time since the closest approach in the west is roughly about 10$^{8}$yrs, consistent 
with the rotational time estimated above. This time scale may linked to the processes of YSCC formation (e.g., 
modes and mechanisms of the induced star formation; feedbacks include radiation/thermal pressures, stellar winds 
and supernova blasts, etc). The existence of a large amount of young (A-type) stars in the nucleus of ARP~24 
(see $\S$3.3) may also be related to the age of interaction in this system. However, such a formation scenario 
still remains speculative, and needs to be confirmed by hydrodynamical modeling (e.g., Struck et al. 2005) and 
numerical simulations of star formation in galaxy mergers (e.g., Struck \& Smith 2003; Cox 2004).

Recent studies have shown that young star clusters tend to form in large complexes instead 
of isolation (Bastian et al. 2005). According to Efremov (2004), these complexes (or so called 'clusters 
of clusters') are often the massive bound clusters formed from a single gas supercloud, and within high 
pressure surroundings. Zhang et al. (2001) made a multi-wavelength study of the mass, age, and 
space distributions of young star clusters in the Antennae galaxy (NGC~4038/39), and found the young 
clusters have a clumpy space distribution and located in regions of high interstellar density. And the 
cluster formation rate ($\Sigma_{CFR}$) is correlated with the interstellar medium (ISM) 
density ($\Sigma_{ISM}$). From the spectroscopic studies of an unusual star complex in NGC~6946 (Larsen 
et al. 2002) using the SAO 6m and Keck 10m telescopes, Efremov et al. (2002) found this complex resembles 
a circular bubble 600pc in diameter with a young super star cluster near its center. And the intensities 
of the emission lines within and around the complex indicate that shock excitation makes a significant 
contribution to the emission from the most energetic region. Larsen (2004b) found a strong correlation 
between cluster age and 'crowding' of the environment, with most of the crowded clusters having young 
ages ($\le$ 10$^{7}$yr). Chen et al. (2005) found a tight group of clusters in the very luminous 
giant H {\sc ii} region: NGC~5461 in spiral galaxy M~101. Bastian et al. (2005) found several young star 
cluster formed in larger groupings/complexes in spiral galaxy M~51. These complexes are all young ($<$10 Myr), 
have sizes between $\sim$85 and $\sim$240 pc, and have masses between 3-30$\times$10$^{4}$M$_{\rm \odot}$, 
similar to the YSCC in ARP~24.

Elmegreen et al. (1993) suggested that the local turbulent velocity dispersion plays an 
important role in setting the mass scale of the star-forming clouds. The larger mass clouds which formed in 
more turbulent conditions, are more resistant to the disruptive effects of young stars, i.e., can sustain a higher 
star-forming efficiency. Clark et al. (2005) investigated the formation of star clusters in unbound giant molecular 
clouds (GMCs), where the cloud tend to form a series of star clusters and disperse quickly, within a rapid star 
formation process ($\sim$10Myr). They also proposed that the clusters that form in the unbound GMCs may be progenitors 
of the OB associations. Zhang et al. (2001) found the youngest star clusters in the Antennae galaxy are associated with 
molecular cloud complexes with characteristic radii of about 1 kpc. Wilson et al. (2003) studied properties of the 
supergiant molecular complexes in the Antennae using CO emission, and suggested that young massive star clusters 
formed from dense cores within the observed supergiant molecular complexes. The star clusters in the YSCC of ARP~24 
may formed in a single GMC simultaneously, since they have similar ages and masses, and reside in a small region 
of $\sim$127pc (see $\S$3.2). Then they may disperse quickly and dissipate to become field population before reaching 
ages of 100Myr according to Wilson et al. (2006). However, this scenario still needs to be confirmed by deep spectroscopic 
observations with high dispersion and high spatial resolution, for measuring the velocity dispersions and obtaining 
chemical abundances of individual clusters in the YSCC (see, e.g., Larsen et al. 2006; Bastian et al. 2006b). There is 
no evidence of merging of star clusters in the center of the YSCC, which was found in many complexes in M~51 
(Bastian et al. 2005).

Alternatively, perhaps the YSCC is a dwarf galaxy which have fallen into ARP~24 previously during a merger event. 
Its blue color ($g-i = -0.84$, see also Fig. 8) is similar to those of irregular galaxies (Fukugita et al. 1995), 
and its absolute magnitude ($\sim -13$) is located at the faint end of the luminosity function of extremely low-luminosity 
galaxies (Blanton et al. 2005). Its metallicity is low (see $\S$3.3), and is comparable to those of local dwarf 
irregular galaxy (e.g., NGC~6822; Lee et al. 2006). Its warm infrared color (high F$_{\rm 24\mu m}$/F$_{\rm 8\mu m}$ 
ratio; see next subsection) is also consistent with that of irregular galaxies (e.g., DDO~053) in the {\it Spitzer} 
Infrared Nearby Galaxies Survey (SINGS, Kennicutt et al. 2003; Dale et al. 2005).

\subsection{PAH and Warm Dust Emission}
From the analysis of SEDs of the YSCs in ARP~24 (see $\S$3.1), we find the YSCC have higher 24$\mu$m emission 
relative to the starlight (3.6$\mu$m) and PAH emission (8$\mu$m) than the nucleus and other infrared bright knots 
(Fig. 9). Regions associated with YSCs (or SSCs) often show 'warm' infrared colors (high 24$\mu$m emission relative 
to that at shorter wavelengths) in several interacting and starburst galaxies. Thuan et al. (1997) 
discovered six super-star clusters (SSCs) with ages $\leq$25 Myr and within a region of $\sim$520 pc diameter 
in the extremely metal poor (Z$\sim$Z$_{\rm \odot}$/41) blue compact dwarf galaxy (BCD) SBS~0335-052. Mid-infrared 
spectrum with {\it ISO} and {\it Spitzer} show no sign of PAH emissions and a relatively flat continuum between 
5 and 20 $\mu$m (Thuan et al. 1999; Houck et al. 2004), which can be explained by the effect of the 
destruction of the PAH molecules by high UV radiation field. Jarrett et al. (2006) found the tidal tail supermassive 
star clusters (with M $\sim$ 10$^{6}$M$_{\rm \odot}$) in the Tadpole Galaxy (UGC~10214) have exceptionally strong 
24$\mu$m emission relative to the starlight, hot dust continuum, and PAH emissions. Elmegreen et al. (2006) studied 
the infrared clumpy emissions in the interacting galaxy pair IC~2163 and NGC~2207, and found the brightest giant 
infrared clump (feature i) which contains two SSCs, numerous other star clusters and associations, several dark 
dust clouds, and supernova 1999ec, have extremely high 24$\mu$m emission relative to 3.6$\mu$m and 8$\mu$m emissions. 
They suggest that this giant clump may formed by gravitational instabilities in the compressed gas of the oval and 
spiral arms. In the western spiral arm of the well-studied nearby spiral galaxy M~51 (NGC~5194), several star-forming 
regions are associated with luminous YSCs and show high F$_{\rm 24\mu m}$/F$_{\rm 8\mu m}$ ratios (e.g., region '05-01' 
shown in Calzetti et al. 2005, associated with the YSC \#180 shown in Larsen 2000; see Fig. 9). Special conditions 
in these infrared warm regions can lead to the formation of extremely young and massive star clusters, with strong 
UV radiation fields which can reduce the PAH feature strength and enhance the warm dust emissions. 

The high F$_{\rm 24\mu m}$/F$_{\rm 8\mu m}$ ratio in the YSCC may be due to the weakness or lack of the PAH emission bands, 
which are caused by low PAH abundance as consequences of PAH destructions by strong UV radiation field and/or other 
mechanisms (e.g., shocks) within this region; or 'PAH-dust competition', i.e., PAH molecules can't be excited by UV 
photons due to the presence of the dust grains (Martin-Hernandez et al. 2006). The elevation of the 24$\mu$m dust 
emission indicates that the VSGs have been heated to a high temperature by their nearby luminous young stars. The 
hardness of the interstellar radiation field may play a major role in the destruction of PAHs in the center 
of star formation regions (e.g., Madden et al. 2006; Bendo et al. 2006), and cause the warm infrared colors. The 
very blue optical color of the YSCC region in ARP~24 (see $\S$3.1) indicates that shocks induced by strong stellar 
winds and/or supernovae explosions could also be a mechanism for destroying PAHs (O'Halloran et al. 2006). Low PAH 
abundance can also originate from low PAH formation rates due to different dust formation processes in a low 
metallicity environment (e.g., Hogg et al. 2005; Engelbracht et al. 2005). Gordon et al. (2006) found the 6.2 and 
7.8+8.6$\mu$m PAH features in M~101 H {\sc ii} regions are weak or absent at a metallicity 12+log(O/H)$\sim$8.0. 
Thus, the low O/H abundance in the region of the YSCC compared with that in the nuclear region (see $\S$3.3) indicates 
that metallicity may play a role in setting infrared colors in ARP~24. However, Bolatto et al. (2006) showed that the 
F$_{\rm 8\mu m}$/F$_{\rm 24\mu m}$ ratios of the regions in the Small Magellanic Cloud (SMC) are regulated primarily by the 
local interstellar radiation field rather than metallicity, and suggested that photo-destruction of PAHs may be 
primarily responsible for the variation of the measured F$_{\rm 8\mu m}$/F$_{\rm 24\mu m}$ ratios in the SMC. Dust can be 
heated by emissions arising from older and cooler stars (e.g., B stars, Peeters et al. 2004), besides ionizing photons 
from young hot stars. The strength of the 24$\mu$m continuum emission from VSGs is more sensitive to the radiation 
energy density and hardness than the 8$\mu$m PAHs (Roussel et al. 2001). Thus, the F$_{\rm 24\mu m}$/F$_{\rm 8\mu m}$ ratio 
also could link to the mean age of the stellar populations responsible for dust excitation. 

Regions associated with YSCs do not always show warm infrared colors. Cannon et al. (2006a) found a lower 
F$_{\rm 24\mu m}$/F$_{\rm 8\mu m}$ ratio and modest warm dust emission of the SSC in the center of nearby dwarf starburst galaxy 
NGC~1705. They suggested that this may due to the dust removal by multi-phase outflow, or evacuation by high UV flux from 
the SSC. The K1 region in ARP~24, which is also associated with a young (t$\sim$3-5 Myr) massive (M$\sim$10$^{5}$M$_{\rm \odot}$) 
star cluster, doesn't show a high F$_{\rm 24\mu m}$/F$_{\rm 8\mu m}$ ratio like that of the YSCC. This indicates the region-to-region 
variations of mid-infrared colors may depend on several physical parameters such as dust composition, dust temperature, and 
PAH molecular abundance. On the other hand, not all the regions with warm infrared colors are associated with optically bright 
YSCs. For example, in NGC~4038/39, the brightest knot in the 15$\mu$m map (Mirabel et al. 1998) has only a faint I-band 
counterpart (object \#80; Whitmore \& Schweizer (1995)), but was found to be associated with a molecular complex (SGMC4-5; 
Wilson et al. 2000). We found that this knot has infrared colors even warmer than that of the YSCC in ARP~24, based on 
archival {\it Spitzer} data (PI: G. Fazio; see also Wang et al. 2004). Its very red near-infrared color 
($J-Ks = 1.86$) indicates that it suffers from heavy dust extinction (Brandl et al. 2005). Thus, we speculate that some 
star clusters in galaxies which are very young and optically faint caused by heavy dust extinctions may also have warm 
infrared colors like the YSCC we found in ARP~24. And this may also caused by the strong UV radiation fields in these 
regions which may destroy the PAH molecules and enhance the VSG emissions at 24$\mu$m. 

Firm conclusions must await a quantitative multi-wavelength analysis of a large, well-defined, and unbiased sample 
of YSCs in galaxies. The precise roles of the PAH destruction and the enhanced 24$\mu$m dust emission in setting the 
F$_{\rm 24\mu m}$/F$_{\rm 8\mu m}$ ratios of the YSCs in ARP~24 could be well-determined by using infrared spectroscopic 
observations. Ground-based mid-infrared imaging and spectroscopy of YSCs with high spatial resolution (subarcsecond; 
e.g., Snijders et al. 2006) will help for studying the spatial distributions and originations of PAH and warm dust 
emissions.

\section{Summary}
In this work, we present a study of young massive clusters in the interacting galaxy ARP~24 using images 
from {\it HST}, {\it Spitzer}, SDSS, {\it GALEX}, H$\alpha$ image, and optical spectroscopy. Our major findings 
are the following: 

(i) From the {\it HST} WFPC2 U and I-band images, we found the brightest infrared knot in ARP~24 
is associated with a complex of young massive star clusters (the YSCC). The ages and masses of the 
star clusters in the YSCC and other three infrared bright knots (K1, K2, K3) were estimated using 
{\it HST} color-magnitude diagrams and comparing with the Starburst99 synthesis models. The clusters 
in the YSCC and K1 region are very young (within ages of around 3-5 Myr) and massive (M$\sim$10$^{5}$
M$_{\rm \odot}$), while the clusters in K2 and K3 regions are relatively less massive (M$\sim$4$\times$10$^{4}$
M$_{\rm \odot}$) but have similar ages ($\sim$ 4-6 Myr) to that in the YSCC and K1 region. The masses 
of the star clusters in the YSCC and K1 are consistent with that of young massive star clusters or 
super star clusters observed in other interacting galaxies.

(ii) Using standard optical emission line ratio diagnostic diagrams and comparing with the 'MAPPINGS III' 
code results, we computed photo-ionization models and estimated the ionization parameter of the YSCC. The 
metallicity of the YSCC was estimated using the emission line ratios, and the star formation rates of the 
YSCC and K1, K2, K3 regions were calculated using monochromatic 24$\mu$m, FUV, and H$\alpha$ line luminosities. 
The SFR density of the region of the YSCC is high and comparable to or even much stronger than the 
definition of a starburst galaxy. The stellar masses for the old stellar population in different regions 
were estimated based on the SDSS photometries and IRAC 3.6$\mu$m luminosities. 

(iii) We speculate that the formation of the YSCs in ARP~24 is triggered by galaxy interaction, and this 
galaxy may formed by a retrograde fly-by encounter indicated by its one-armed appearance and fan-like 
structure. The clusters in the YSCC may formed in a single giant molecular cloud simultaneously since they 
have similar ages and masses, and reside in a small region of $\sim$127pc.

(iv) From the ultraviolet to mid-infrared spectral energy distributions (SEDs), we found that the region 
of the YSCC is relatively bluer in optical and has higher 24$\mu$m dust emission relative to the 
emission at 8$\mu$m. We speculate that this is primarily due to strong UV radiation field (or shocks) within 
this region which may destroy the PAHs and enhance the VSGs emission at 24$\mu$m. However, firm conclusions 
must await a well-defined sample and upcoming infrared spectroscopic observations.

\acknowledgments
We are gratefully to the referee: Dr C. Struck for his very careful review of this paper and many constructive 
advice and very useful suggestions. We thank R. Kennicutt, R. de Grijs, J. Wang, Z.-G. Deng for advice and helpful 
discussions, and Z. Wang, J.-S. Huang, C.-N. Hao, J.-L. Wang, F.-S. Liu for their capable help and assistance 
throughout the process of {\it Spitzer} data reductions. Special thanks go to the staff at Xinglong observatory 
of National Astronomical Observatories, Chinese Academy of Sciences (NAOC) for their instrumental and observational 
helps. This project is supported by NSFC No.10273012, No.10333060, No.10473013, No.10373008. This work is based 
in part on observations made with the NASA/ESA {\it Hubble Space Telescope}, obtained from the Data Archive at 
the Space Telescope Science Institute (STScI). STScI is operated by the Association of Universities for Research 
in Astronomy, Inc., under NASA contract NAS5-26555. The observations of ARP~24 are associated with program IDs 
8599 (T. B{\"o}ker) and 9124 (R. A. Windhorst). This work is based in part on observations made with the 
{\it Spitzer Space Telescope}, which is operated by the Jet Propulsion Laboratory, California Institute of 
Technology under NASA contract 1407. The {\it Galaxy Evolution Explorer} ({\it GALEX}) is a NASA Small Explorer. 
The mission was developed in cooperation with the Centre National d'Etudes Spatiales of France and the Korean 
Ministry of Science and Technology. Funding for the SDSS and SDSS-II has been provided by the Alfred P. Sloan 
Foundation, the Participating Institutions, the National Science Foundation, the U.S. Department of Energy, the 
National Aeronautics and Space Administration, the Japanese Monbukagakusho, the Max Planck Society, and the Higher 
Education Funding Council for England. The SDSS Web Site is http://www.sdss.org/. The SDSS is managed by the 
Astrophysical Research Consortium for the Participating Institutions. The Participating Institutions are the 
American Museum of Natural History, Astrophysical Institute Potsdam, University of Basel, Cambridge University, 
Case Western Reserve University, University of Chicago, Drexel University, Fermilab, the Institute for Advanced 
Study, the Japan Participation Group, Johns Hopkins University, the Joint Institute for Nuclear Astrophysics, 
the Kavli Institute for Particle Astrophysics and Cosmology, the Korean Scientist Group, the Chinese Academy 
of Sciences (LAMOST), Los Alamos National Laboratory, the Max-Planck-Institute for Astronomy (MPIA), the 
Max-Planck-Institute for Astrophysics (MPA), New Mexico State University, Ohio State University, University 
of Pittsburgh, University of Portsmouth, Princeton University, the United States Naval Observatory, and the 
University of Washington. 

{\it Facilities:} \facility{HST (WFPC2)}, \facility{Spitzer (IRAC, MIPS)}, \facility{Sloan}, \facility{GALEX}, 
\facility{Beijing:2.16m (BFOSC, OMR)}.


\clearpage

%
\begin{figure}
\center
\includegraphics[angle=0,scale=.8]{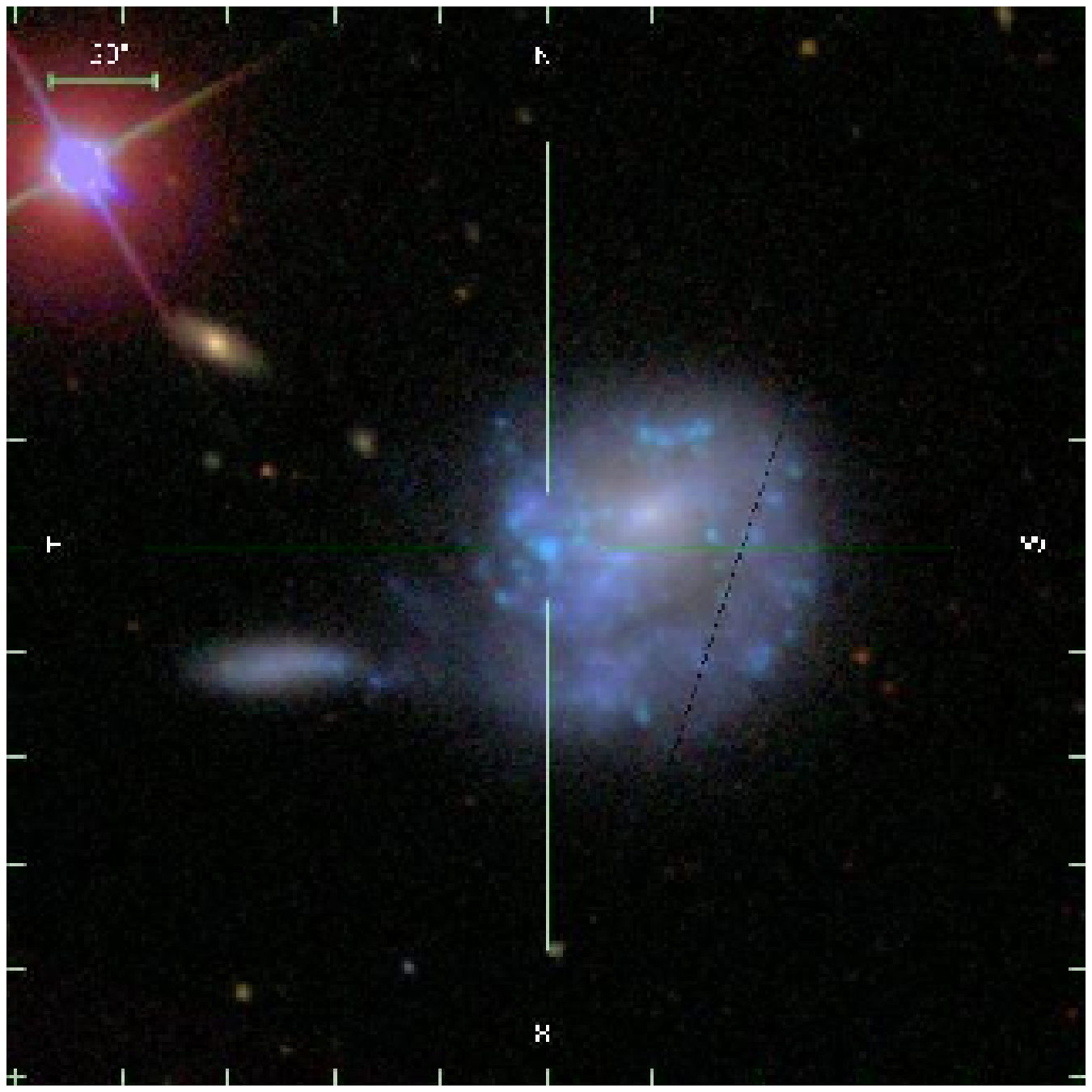}
\caption{Three color image of ARP~24 derived from the SDSS data archive. North is up, 
east is to the left as denoted by the cross hair, and the center is the region of 
the YSCC in this galaxy (see the text). \label{fig1}}
\end{figure}
\begin{figure}
\center
\includegraphics[angle=0,scale=.6]{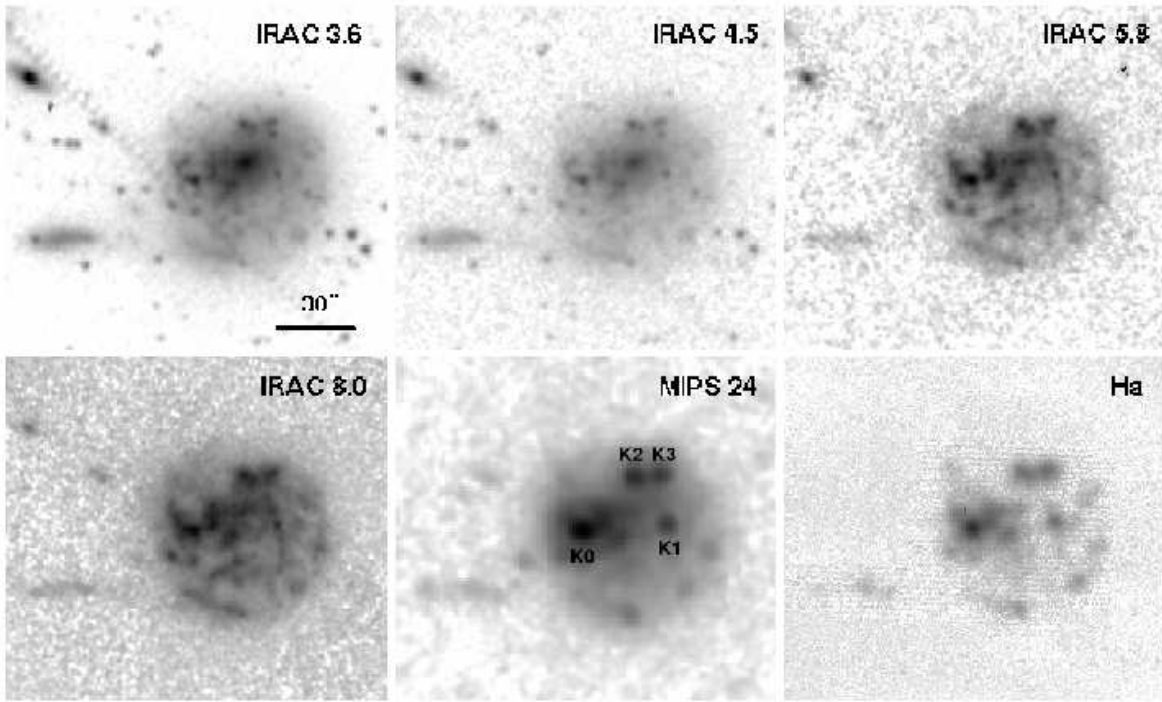}
\caption{{\it Spitzer} IRAC 3.6, 4.5, 5.8, 8$\mu$m and MIPS 24$\mu$m bands images of ARP~24. The 
continuum-subtracted H$\alpha$ image was also plotted to emphasize the good correspondence of the 
peaks between H$\alpha$ and 24$\mu$m emissions. K0, K1, K2, K3 denote the four infrared bright 
knots in ARP~24 selected from the 24$\mu$m image as we described in the text (see $\S$3.1). \label{fig2}}
\end{figure}
\begin{figure}
\center
\includegraphics[angle=90,scale=.37]{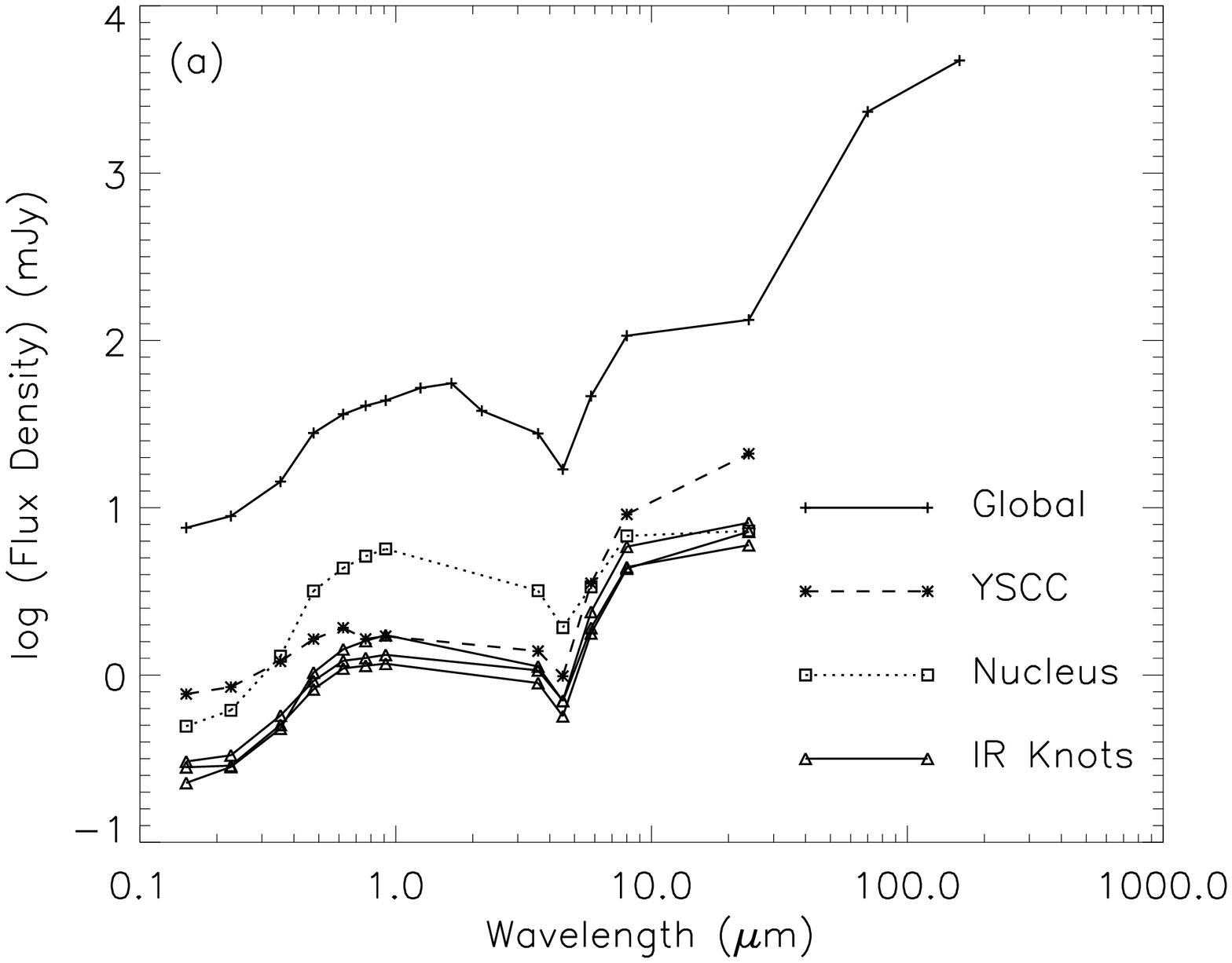}
\includegraphics[angle=90,scale=.37]{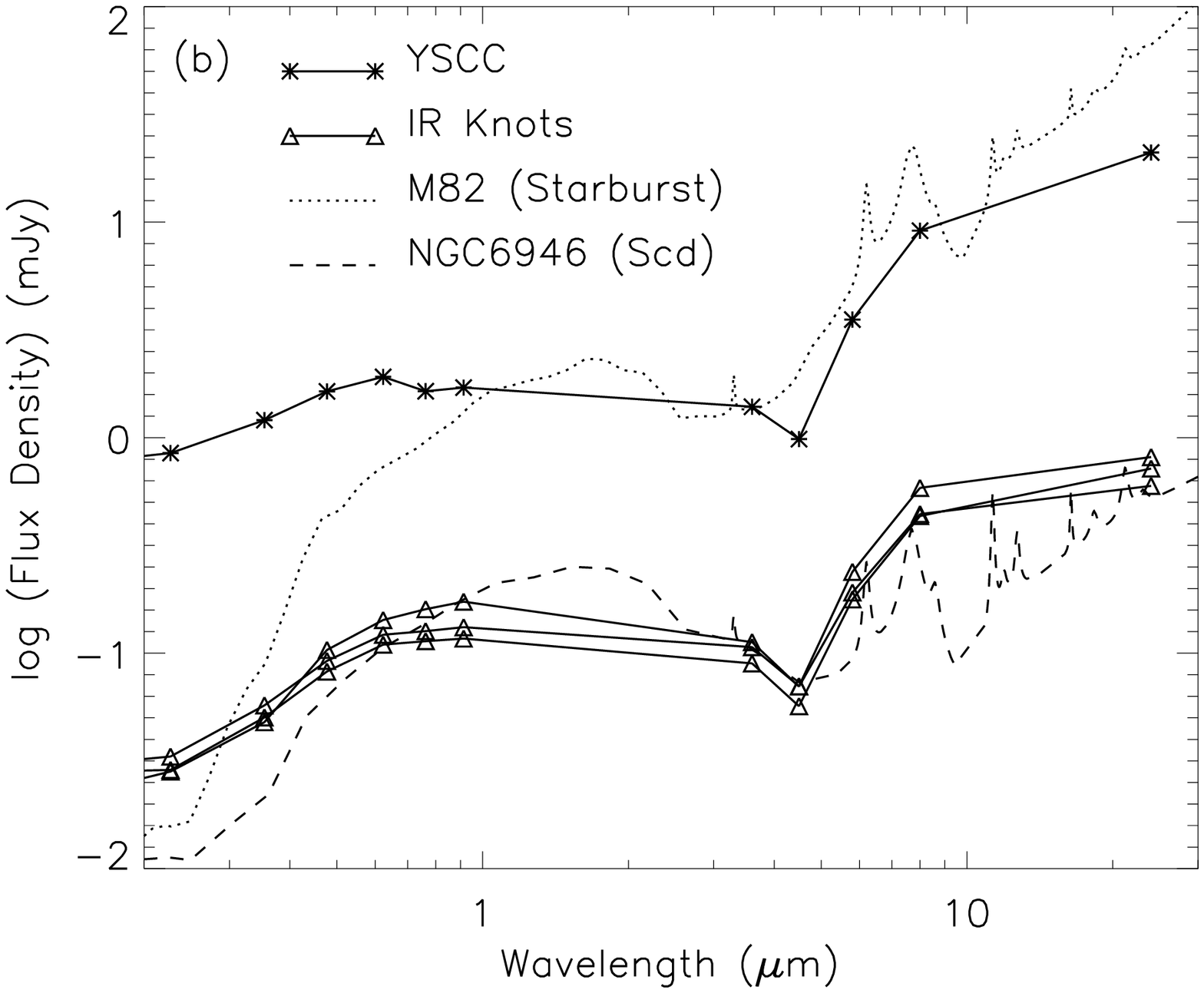}
\caption{(a) Spectral Energy Distributions (SEDs) of ARP~24, its nucleus, the YSCC, and other three infrared 
bright knots (K1, K2, K3; IR Knots); (b) SEDs of the YSCC and IR Knots, compared with the empirical SED 
templates generated using the GRASIL code and normalized to the average of SDSS z-band and IRAC 3.6$\mu$m flux. The 
YSCC is similar to the prototype starburst galaxy M~82 (dotted line), and the IR Knots are comparable to late-type 
Scd galaxy NGC~6946 (dashed line). Note the flux densities of the IR Knots have been divided by a factor 10 (1 dex) 
for clarity. \label{fig3}}
\end{figure}
\begin{figure}
\center
\includegraphics[angle=0,scale=.35]{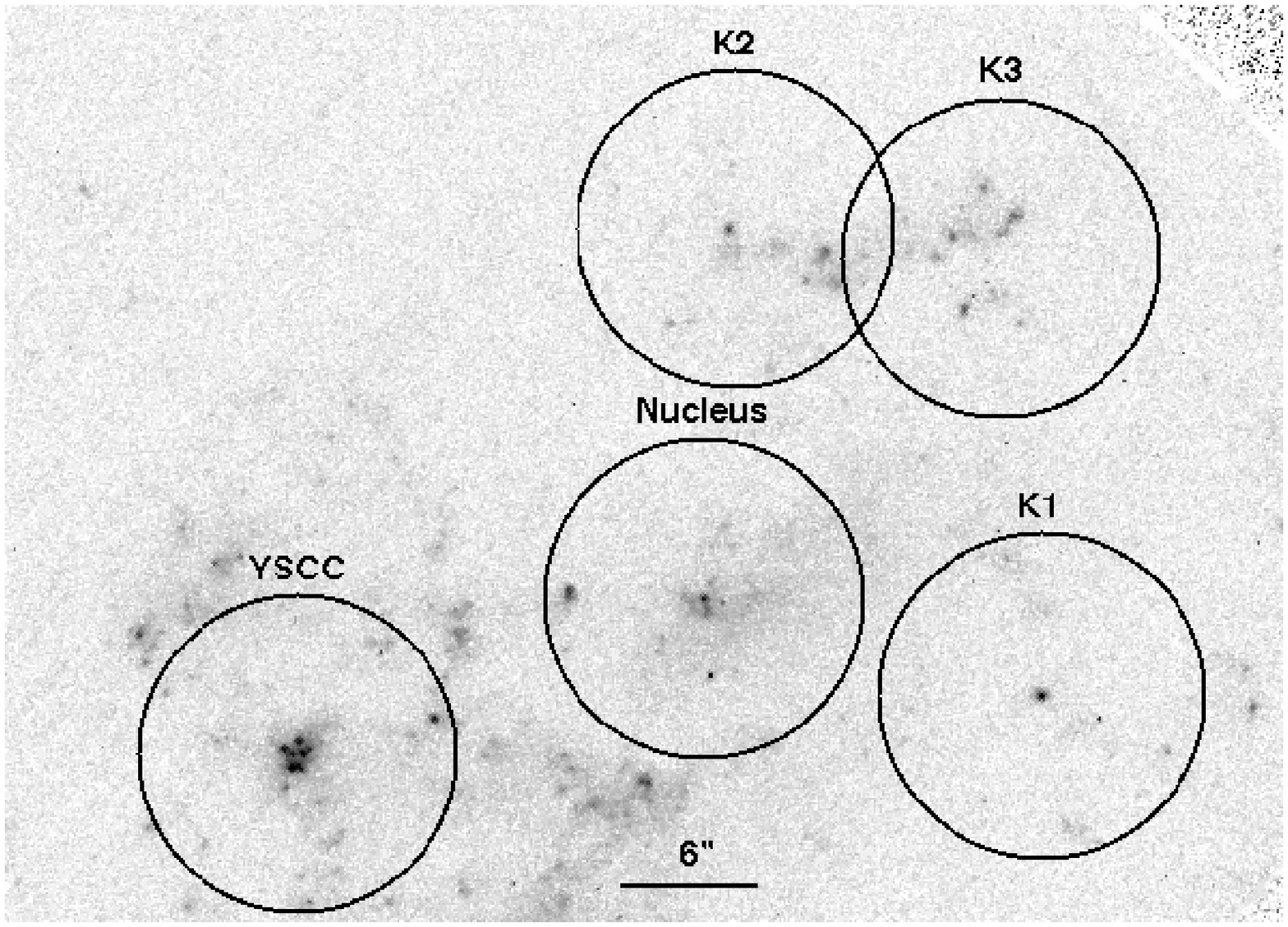}
\includegraphics[angle=0,scale=.35]{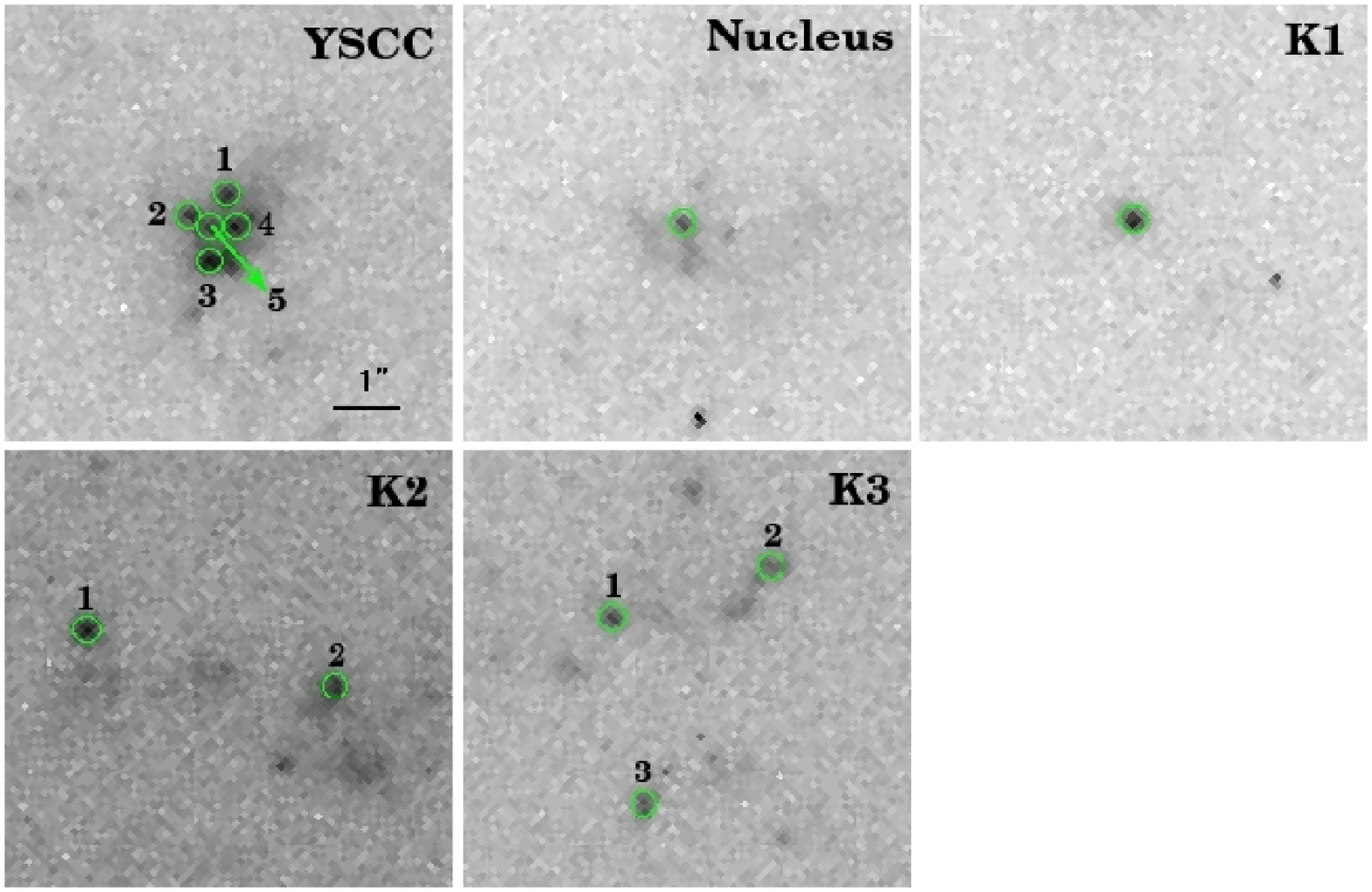}
\caption{{\it HST} WFPC2 F300W image of star clusters in the young massive star cluster 
complex (YSCC), the nucleus and three infrared bright knots (K1, K2, K3) in ARP~24, north is up, 
east is to the left. The black circles shown in the left image correspond to the 6$''$ radius apertures 
used in the SED studies (see $\S$3.2), and the green ones shown in the right image correspond to 0.2$''$ 
radius apertures on the F300W image. \label{fig4}}
\end{figure}
\begin{figure}
\center
\includegraphics[angle=90,scale=.7]{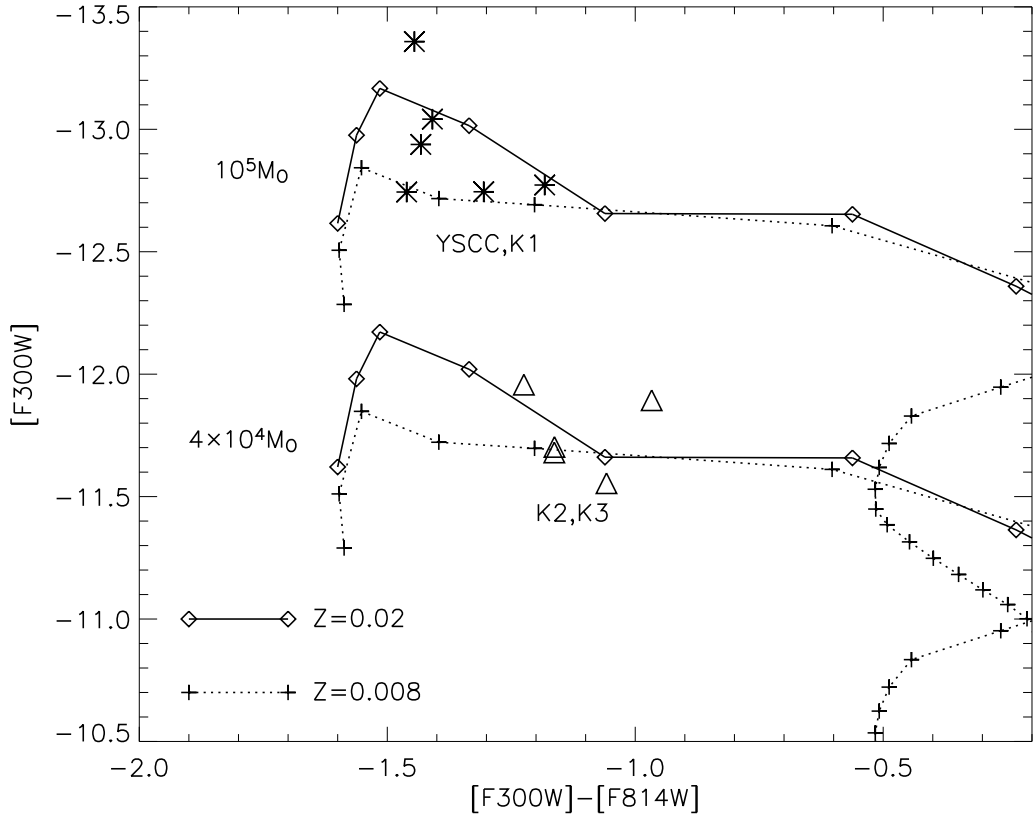}
\caption{{\it HST} WFPC2 color-magnitude diagrams (CMD) of star clusters in the YSCC and 
three infrared bright knots (K1, K2, K3) in ARP~24. Starburst99 instantaneous models with Salpeter 
initial mass function (IMF, $\alpha_{\rm IMF} = 2.35$) between 0.1 and 120 M$_{\rm \rm \odot}$ and 
metallicities Z=0.02 \& 0.008 are also shown in this figure. Evolutionary lines are drawn for 
cluster masses of 10$^{5}$ and 4$\times$10$^{4}$ M$_{\rm \odot}$. Model ages along the evolutionary 
lines are indicated by {\bf $\diamond$} (Z=0.02) and {\bf +} (Z=0.008) placed in t=1 Myr intervals 
and begin from 1 Myr on the left. \label{fig5}}
\end{figure}
\begin{figure}
\center
\includegraphics[angle=90,scale=.7]{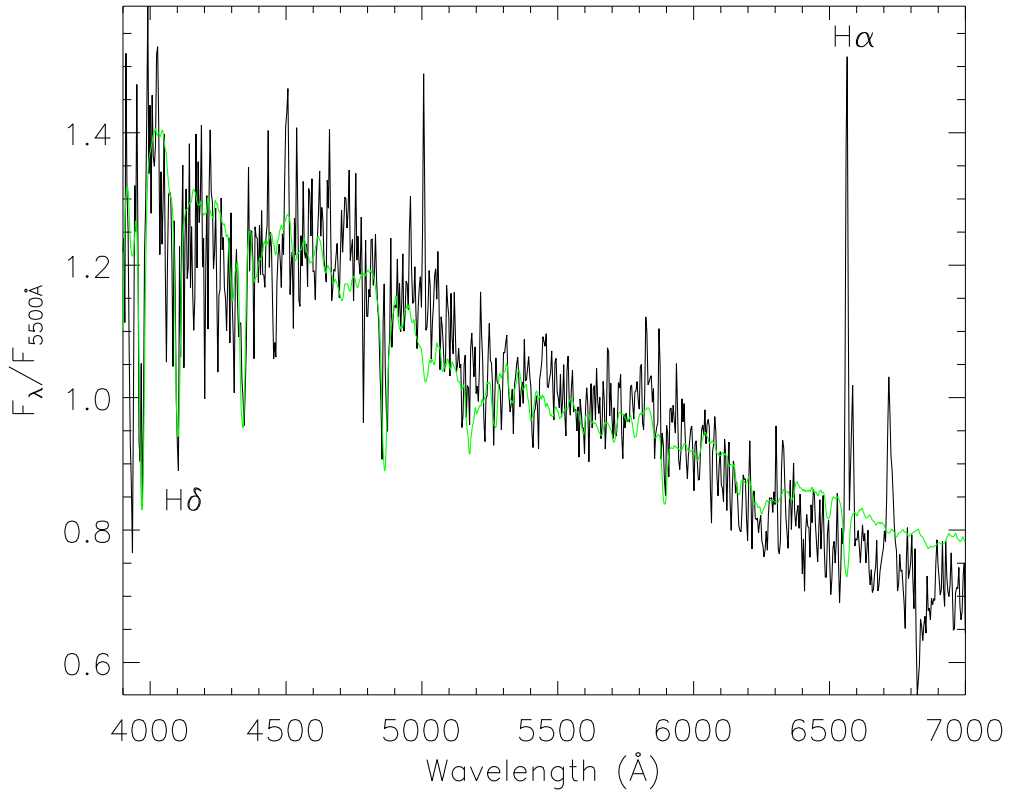}
\caption{Optical spectrum of the nucleus in ARP~24 (plotted in black; normalized to L$_{\rm \lambda}$(5500\AA)=1), 
which indicates a mixture of populations of different ages. Strong H$\delta$ absorption indicates the existence 
of a large number of evolved A-type stars (age $\sim$ 10$^{8}$yr), and strong H$\alpha$ emission traces ongoing 
star formation. A template spectrum corresponds to an instantaneous-burst model with a young population of 100 Myr 
and metallicity Z=0.02 plus an old population of 11 Gyr was also plotted for comparison (in green; derived from 
Bruzual \& Charlot 2003; Note the template has been convolved to match the observed spectrum). \label{fig6}}
\end{figure}
\begin{figure}
\center
\includegraphics[angle=0,scale=.8]{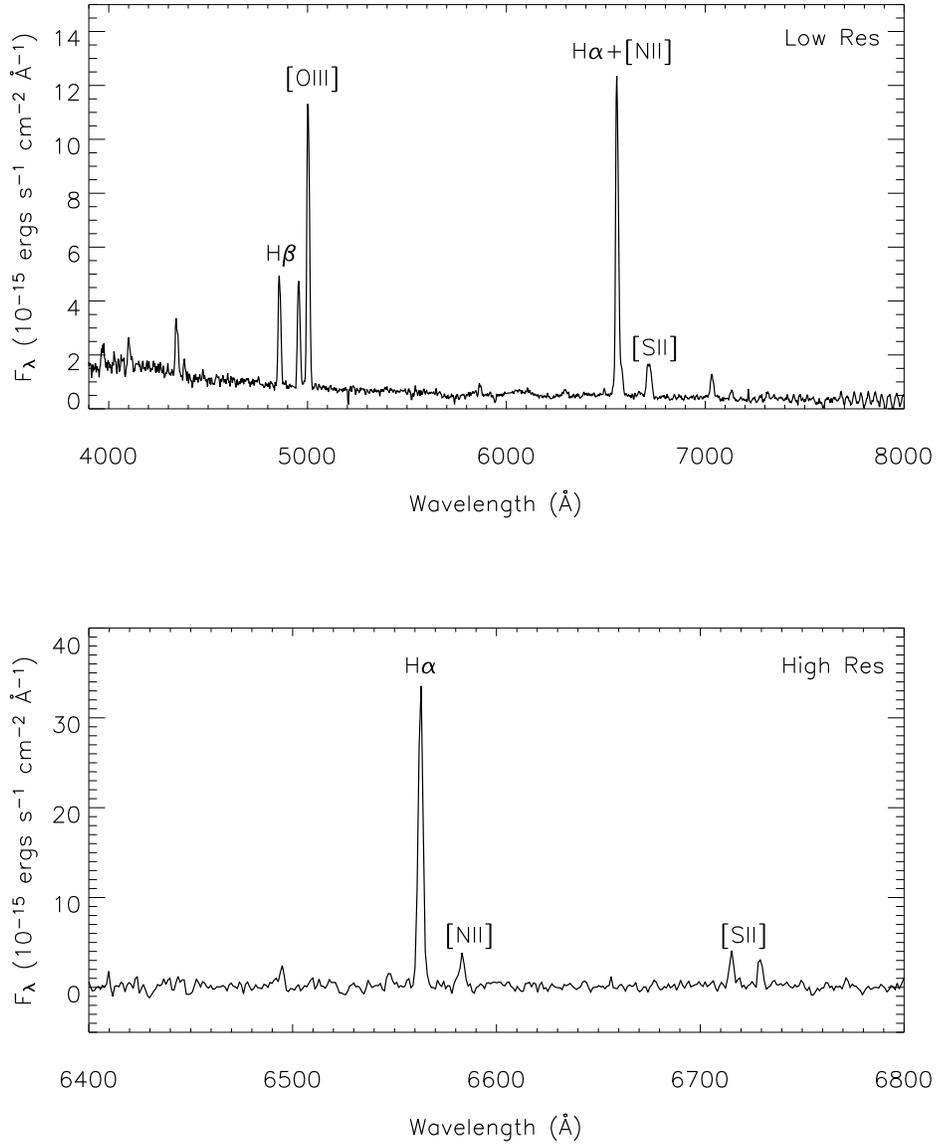}
\caption{Low (top) and high (bottom) resolution optical spectra of the YSCC in ARP~24. \label{fig7}}
\end{figure}
\begin{figure}
\center
\includegraphics[angle=0,scale=.7]{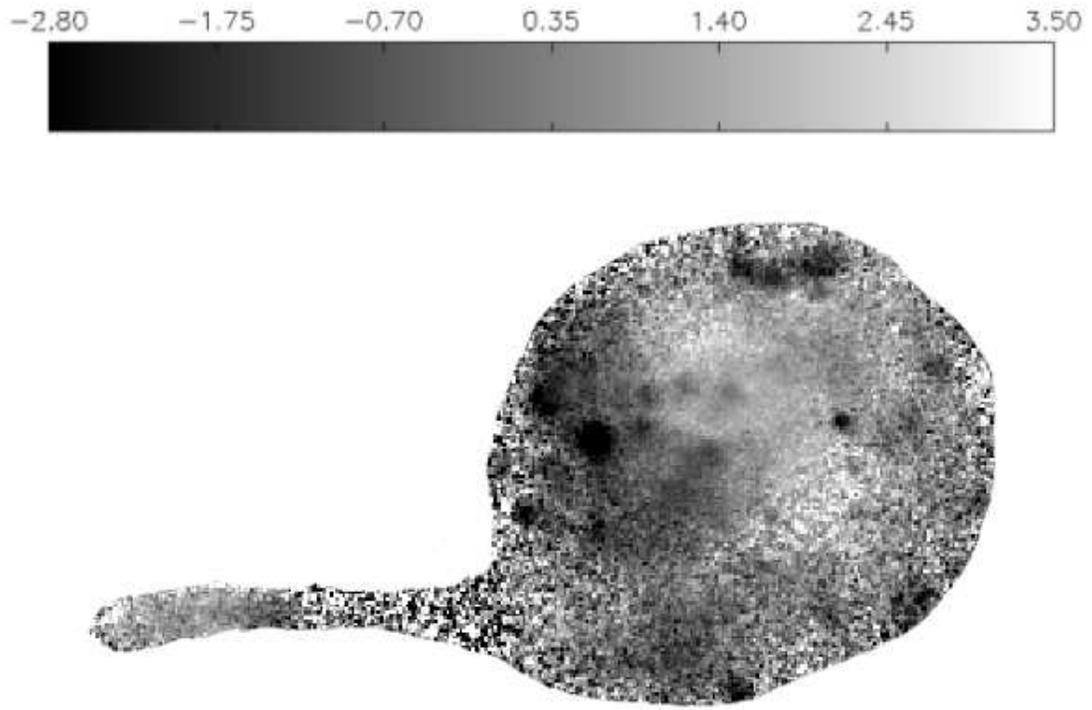}
\caption{SDSS g-i color map of ARP~24. There is a strong asymmetry through the galactic disk: the regions near the 
YSCC are relatively bluer than that near the nucleus. This asymmetry may be due to the differences in star 
formation, stellar populations, gas/dust contents of different regions caused by galaxy interactions.\label{fig8}}
\end{figure}
\begin{figure}
\center
\includegraphics[angle=90,scale=.7]{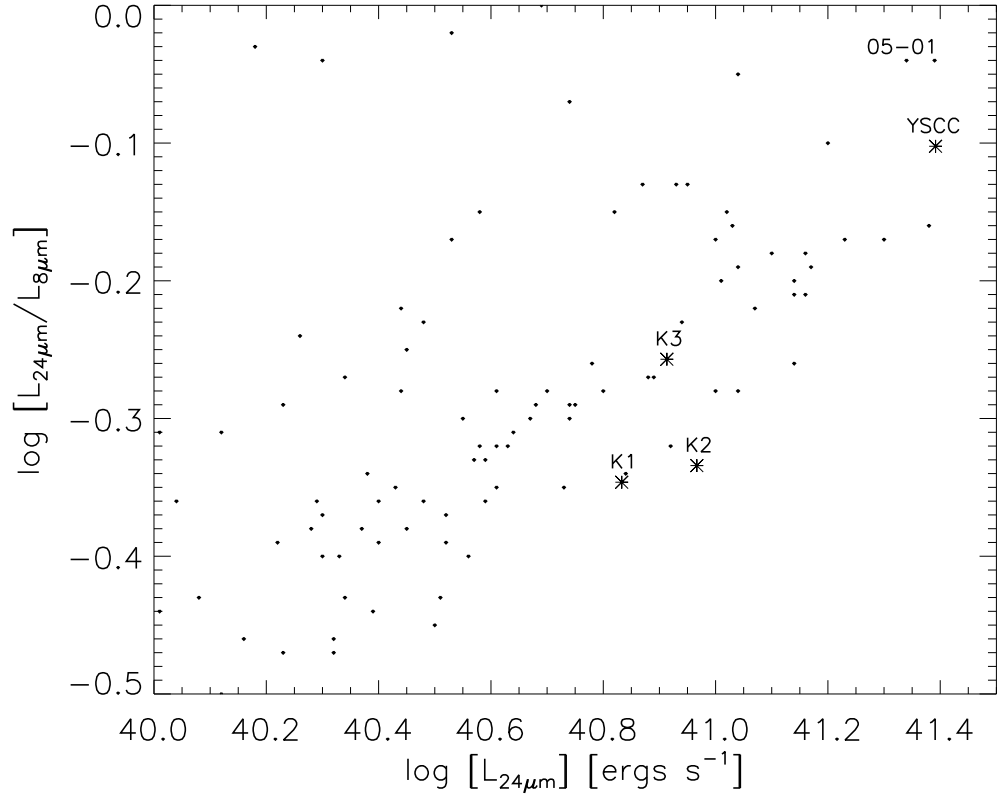}
\caption{Mid-infrared magnitude-color diagram of the YSCC and three infrared bright knots (stars) in ARP~24. Compared 
with that of H {\sc ii} regions (dots) in M~51 (Calzetti et al. 2005). \label{fig9}}
\end{figure}

\clearpage

%
\begin{table}[]
  \caption[]{Spectral Energy Distributions}
  \label{Tab:1}
  \begin{center}\begin{tabular}{c|cccccccccccccc}
  \hline
  \hline
Regions & YSCC & Nucleus & K1 & K2 & K3 & Total\\
\hline
RA & 10:54:37.9 & 10:54:35.5 & 10:54:34.1 & 10:54:35.5 & 10:54:34.4 & 10:54:35.5\\
DEC & +56:59:20 & +56:59:26 & +56:59:23 & +56:59:41 & +56:59:41 & +56:59:26\\
\hline
FUV & 0.77 & 0.49 & 0.23 & 0.30 & 0.28 & 7.59\\                     
NUV & 0.85 & 0.62 & 0.28 & 0.33 & 0.29 & 8.91\\
\hline
u & 1.21 & 1.30 & 0.48 & 0.57 & 0.50 & 14.30\\                        
g & 1.64 & 3.18 & 1.04 & 0.92 & 0.82 & 27.97\\
r & 1.92 & 4.36 & 1.43 & 1.22 & 1.10 & 36.21\\
i & 1.64 & 5.13 & 1.60 & 1.27 & 1.14 & 40.70\\
z & 1.71 & 5.67 & 1.73 & 1.32 & 1.17 & 43.76\\
\hline
J & \ldots & \ldots & \ldots & \ldots & \ldots & 52.0\\
H &  \ldots & \ldots & \ldots & \ldots & \ldots & 55.4\\
Ks & \ldots & \ldots & \ldots & \ldots & \ldots & 38.0\\
\hline
3.6$\mu$m & 1.39 & 3.19 & 1.13 & 1.07 & 0.90 & 27.76\\ 
4.5$\mu$m & 0.99 & 1.92 & 0.70 & 0.70 & 0.57 & 16.96\\
5.8$\mu$m & 3.53 & 3.37 & 1.91 & 2.38 & 1.78 & 46.42\\
8$\mu$m  &  9.13 & 6.79 & 4.42 & 5.85 & 4.33 & 106.66\\
\hline
24$\mu$m & 21.63 & 7.27 & 5.97 & 8.13 & 7.19 & 132.67\\
70$\mu$m & \ldots & \ldots & \ldots & \ldots & \ldots & 2326.96\\
160$\mu$m & \ldots & \ldots & \ldots & \ldots & \ldots & 4710.42\\
\hline
H$\alpha$ & 40.26 & \ldots & 39.57 & 39.72 & 39.70 & 40.66\\
\hline
\hline
\end{tabular}\end{center}
\begin{list}{}{}
\item[]Notes.$\--$The fluxes of {\it GALEX} two-bands (FUV \& NUV), SDSS five-bands ($u,g,r,i,z$), 
2MASS three-bands (J, H, Ks), {\it Spitzer} IRAC four-bands (3.6, 4.5, 5.8, 8$\mu$m) and MIPS three-bands 
(24, 70, 160$\mu$m) are in units of mJy. The H$\alpha$ line luminosities are in units of log (ergs s$^{-1}$), 
after correcting the contamination of [NII] lines based on the [NII]/H$\alpha$ ratio derived from the 
spectroscopy (see $\S$3.3). 
\end{list}
\end{table}

%
\begin{table}[]
  \caption[]{{\it HST} WFPC2 F300W and F814W fluxes and magnitudes}
  \label{Tab:2}
  \begin{center}\begin{tabular}{cc|cc|cc|cc|c}
  \hline
  \hline
Clusters & ~ & RA & DEC & F$_{\rm F300W}$ & F$_{\rm F814W}$ & m$_{\rm F300W}$ & m$_{\rm F814W}$ & m$_{\rm F300W}$-m$_{\rm F814W}$\\
~ & ~ & (10:54) & (+58:59) & ~ & ~ & ~ & ~ & ~\\
  \hline
YSCC & ...1 & 37.87 & 21.9 & 59.90 & 6.44 & -12.77 & -11.59 & -1.18\\
~ & ...2 & 37.95 & 21.6 & 58.35 & 4.86 & -12.74 & -11.28 & -1.46\\
~ & ...3 & 37.90 & 20.9 & 102.73 & 8.67 & -13.36 & -11.91 & -1.45\\
~ & ...4 & 37.85 & 21.4 & 76.74 & 6.70 & -13.04 & -11.63 & -1.41\\
~ & ...5 & 37.90 & 21.4 & 58.40 & 5.61 & -12.75 & -11.44 & -1.31\\
\hline
Nucleus & ~ & 35.54 & 26.4 & 41.02 & 26.42 & -12.36 & -13.12 & 0.76\\
\hline
K1      & ~ & 34.12 & 22.8 & 69.76 & 5.96 & -12.94 & -11.51 & -1.43\\
\hline
K2 & ...1 & 35.68 & 41.9 & 28.27 & 2.92 & -11.96 & -10.73 & -1.23\\
~ & ...2 & 35.17 & 40.8 & 21.90 & 2.39 & -11.68 & -10.52 & -1.16\\
\hline
K3 & ...1 & 34.52 & 42.0 & 26.65 & 3.49 & -11.89 & -10.93 & -0.96\\
~ & ...2 & 34.25 & 42.5 & 19.51 & 2.35 & -11.55 & -10.50 & -1.05\\
~ & ...3 & 34.44 & 38.9 & 22.36 & 2.44 & -11.70 & -10.54 & -1.16\\
\hline
\hline
  \end{tabular}\end{center}
\begin{list}{}{}
\item[]Notes.$\--$The fluxes (F$_{\rm F300W}$ \& F$_{\rm F814W}$) are in units of 10$^{-18}$ergs$^{-1}$cm$^{-2}$\AA$^{-1}$, 
and the magnitudes (m$_{\rm F300W}$ \& m$_{\rm F814W}$) are in VEGA systems.
\end{list}
\end{table}

%
\begin{table}[]
  \caption[]{Optical line fluxes of the YSCC}
  \label{Tab:3}
  \begin{center}\begin{tabular}{lccc}
  \hline
  \hline
Line & $\lambda_{0}$ & Flux & Spectrum\\
(\AA) & (\AA) & (10$^{-15}$ergs cm$^{2}$s$^{-1}$) & ~\\
\hline
H$\beta$ $\lambda$4861 & 4895 & 67.0$\pm$1.0 & Low Res\\
$\lbrack$OIII$\rbrack$ $\lambda$4959~~~~~~ & 4994 & 63.4$\pm$1.0 & Low Res\\
$\lbrack$OIII$\rbrack$ $\lambda$5007 & 5042 & 176.0$\pm$2.0 & Low Res\\
H$\alpha$ $\lambda$6563 & 6609 & 210.0$\pm$2.0 & High Res\\ 
$\lbrack$NII$\rbrack$ $\lambda$6583 & 6629 & 25.5$\pm$1.0 & High Res\\
$\lbrack$SII$\rbrack$ $\lambda$6718 & 6765 & 21.8$\pm$1.0 & High Res\\
$\lbrack$SII$\rbrack$ $\lambda$6733 & 6780 & 18.8$\pm$1.0 & High Res\\
\hline
\hline
\end{tabular}\end{center}
\begin{list}{}{}
\item[]Notes.$\--$The line fluxes of the high resolution spectrum (High Res) has been scaled to 
the low resolution one (Low Res) using H$\alpha$ line fluxes. This correction will not affect much 
on the line ratios used in $\S$3.3.
\end{list}
\end{table}

\end{document}